\documentclass{IEEEcsmag}
\usepackage{booktabs} % For formal tables
\usepackage{subcaption}
\usepackage{balance}
\usepackage{graphics}
\usepackage{hyperref}
\usepackage{todonotes}

\begin{document}
\title{What if we have Meta GPT? Content Singularity and Human-Metaverse Interaction in AIGC Era}
% \titlenote{Produces the permission block, and
% \subtitle{Extended Abstract}
% \subtitlenote{The full version of the author's guide is available as
%   \texttt{acmart.pdf} document}

\author{\protect Lik-Hang Lee}
\affil{%
 Hong Kong Polytechnic University, Hong Kong SAR
}

% \email{likhang.lee@kaist.ac.kr}

\author{\protect Pengyuan Zhou}
\affil{%
 University of Science and Technology of China, China
}

\author{\protect Chaoning Zhang}
\affil{%
 Kyung Hee University, South Korea
}

\author{\protect Simo Hosio}
\affil{%
 University of Oulu, Finland
}

% \author{Ahmad Alhilal}
% \affil{%
%    Hong Kong University of Science and Technology, Hong Kong
% }
% % \email{aalhilal@ust.hk}

% \author{Carlos Bermejo Fernández}
% \affil{%
% Hong Kong University of Science and Technology, Hong Kong
% }
% % \email{cbf@ust.hk}

% \author{Pan Hui}
% \affil{%
% Hong Kong University of Science and Technology, Hong Kong \\
% University of Helsinki, Finland
% }
% \affiliation{%
%   \institution{University of Helsinki, Finland}
%   %\city{Helsinki}
%     \country{}
% }
% \email{panhui@ust.hk}

% % The default list of authors is too long for headers}
% \renewcommand{\shortauthors}{T. Braud et al.}

% https://www.lbbonline.com/news/the-metaverse-did-it-live-up-to-the-hype-in-2022
% target: 11 pages including everything. (around 4900 words)
\begin{abstract}

The global metaverse development is facing a ``cooldown moment", while the academia and industry attention moves drastically from the Metaverse to AI Generated Content (AIGC) in 2023. Nonetheless, the current discussion rarely considers the connection between AIGCs and the Metaverse. We can imagine the Metaverse, i.e., immersive cyberspace, is the black void of space, and AIGCs can simultaneously offer content and facilitate diverse user needs. As such, this article argues that AIGCs can be a vital technological enabler for the Metaverse. 
The article first provides a retrospect of the major pitfall of the metaverse applications in 2022. Second, we discuss from a user-centric perspective how the metaverse development will accelerate with AIGCs. Next, the article conjectures future scenarios concatenating the Metaverse and AIGCs. Accordingly, we advocate for an AI-Generated Metaverse (AIGM) framework for energizing the creation of metaverse content in the AIGC era. 

\end{abstract}

% <concept_id>10011007.10010940.10010941.10010949</concept_id>
% <concept_desc>Software and its engineering~Operating systems</concept_desc>
% <concept_id>10010147.10010371.10010387.10010392</concept_id>
% <concept_desc>Computing methodologies~Mixed / augmented reality</concept_desc>
% <concept_id>10003120.10003138.10003140</concept_id>
% <concept_desc>Human-centered computing~Ubiquitous and mobile computing systems and tools</concept_desc>

% \ccsdesc[500]{Software and its engineering~Operating systems}
% \ccsdesc[500]{Computing methodologies~Mixed / augmented reality}
% \ccsdesc[500]{Human-centered computing~Ubiquitous and mobile computing systems and tools}

%\keywords{Extended Reality, Operating Systems}

\maketitle

\section{Retrospect: Experimental Metaverse}

We have witnessed a surge of investment and rigorous discussion regarding the Metaverse since 2021. Many believe a fully realized metaverse is not far off, so tech firms, e.g., Meta, Niantic, Roblox, Sandbox, just to name a few, have started creating their immersive cyberspaces with diversified visions and business agendas. After the metaverse heat wave in 2022, all of us are still as vague about what the Metaverse is. At the same time, the hype surrounding the metaverse shows a sign of slowing down %a negative sign of decelerating 
anytime soon, primarily due to multiple metrics reflecting constantly low numbers of active daily users, the decreasing volume of projects, and the high uncertainty of return on investment. 

When the tech giants dipped their toes into the experimentation pool in 2022, they brought a few playful tasks to their self-defined virtual venues, giving users something to do.
The fascinating difficulty is that the metaverse is already fundamentally split among the forwarding-thinking firms establishing their metaverse realm. %As such, these 
Due to limited time and resources, these firms tried hard to resolve technical issues that shape their immersive cyberspace, such as developing efficient infrastructure that supports unlimited numbers of users in the same virtual venues or offering a decentralized transaction ecosystem driven by blockchain technology. 

Nonetheless, content development is delegated to third parties and thus goes beyond the firms' core concerns. Tech firms commonly leave content creation to the designers and creators, having an unattainable hope that designers and creators can fill up the rest of the metaverse. As a result, one can argue the current virtual spaces have become aimless, primarily caused by the lack of content and, therefore, activities, while users cannot find good reasons to spend time at such venues daily. Moreover, the experimental metaverses of 2022 often neglect usability issues leading to user experiences far from satisfactory. A prominent example is that first-time users struggle to understand the interaction techniques with their avatars in 3D virtual environments. Even worse, after hours of practice, these unskillful users can still not master such interaction techniques, causing poor usability entirely. Without addressing the gaps in content and usability, the firms' ambition exceeds what is practically feasible. Their ambition refers to %It is expected that 
the massive uses of the Metaverse, i.e., the immersive cyberspace~\cite{Lee2022WhatIT}. The core values surrounding the users are not there yet to make the Metaverse a reality. 

We can briefly look back at the transition from the static web (Web 1.0) to its interactive counterpart (Web 2.0) in the 2D-UIs era, characterized by the empowerment of content creation.
Among the static webpages in Web 1.0, limited people with relevant skills can publish information online. At the same time, users can only read the information but have no way to make a two-way interaction. Accordingly, Web 2.0, also known as social networks (SNS), offers participatory and dynamic methods and empowers two-way user interaction, i.e., reading and writing information in 2D UIs. The critical transition from Web 1.0 to 2.0 is that users, regardless of their technology literacy, can freely contribute content on SNS, such as text and images, and then put the content online. We must note that we are good at writing a message on a (soft-)keyboard and taking photos or videos with cameras. Also, most 2D UIs follow certain design paradigms, requiring only simple yet intuitive interactions like clicks, taps, swipes, drags, etc., to accomplish new content creation. 

In contrast, although the metaverse supposedly allows everyone to access many different virtual worlds, three unprecedented barriers arise. 
First, the current users have extensive experience with 2D UIs but not their 3D counterparts. As the entire experiences proceed with 3D UIs, the users in the Metaverse %(perhaps Web 3.0) 
have to deal with unfamiliar virtual worlds with increasing complexity. More importantly, 3D objects do not explicitly show user interaction cues. As the Metaverse claims to be digital twins of our physical environment~\cite{lee2021needs}, a user encounters a virtual chair and employs analogies between the virtual and physical worlds. A simple question could be: Can the user's virtual hands lift or push the chair? As such, users, in general, may not be aware of how virtual objects interact in the Metaverse and thus rely on educated guesses and trial-and-error approaches. 
The above question can be generalized into sub-questions, including but not limited to: What are the available interaction techniques? When to activate the user-object interaction? How does the user understand the related functions mapped to the object? How can we manage the user expectation after a particular click? Which visual and audio effects impact the user's task performance? 

\begin{figure*}[hpt]
    \centering
    \includegraphics[width=15.5cm]{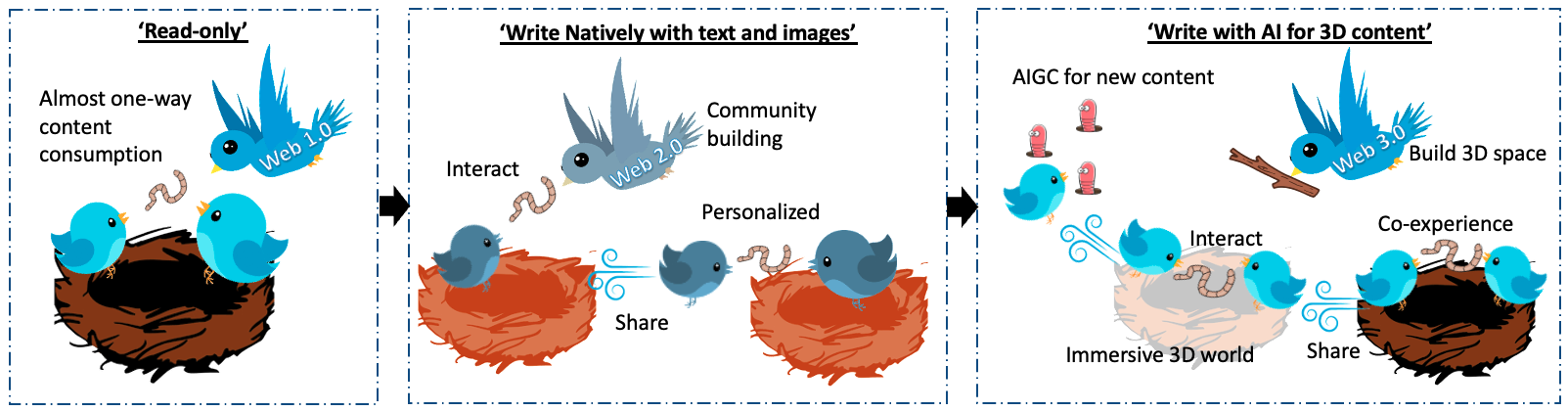}
    \caption{AIGCs can prevent us from falling into another `Web 1.0' in the metaverse era -- the layman end-users suffer from the missing capability of creating unique content. We are natively skilful at texting and photo-taking in social networks but not editing 3D content in virtual 3D spaces. AIGCs may serve as a saviour to enable the general users freely express themselves, while owners of the platforms or virtual spaces can still delegate the content creation tasks to the peer users.}
    \label{fig:issue}
\end{figure*}

Second, the current interaction techniques allow users to manipulate a virtual object, such as selecting, rotating, translating, etc. Still, user efforts in object manipulation are a big concern. Commercial input hardware for headsets (e.g., controllers or joysticks) or even hand gestural inputs are barely sufficient for simple point-and-select operations on 2D UIs in virtual environments~\cite{ubipoint} but largely insufficient for 3D models, especially those with irregular shape causing intolerably long editing time and high dissimilarity with the intended shape \cite{3DeformR}.  
Therefore, users with the current techniques, primarily point-and-select or drag-and-drop, can only manipulate objects with low granularity. However, content creation involves careful manipulation of a 3D object, i.e., modifying the vertex positions in great detail.
Even though nowadays users engage in an immersive 3D environment, most can only create 2D texts and select some standard 3D objects from an asset library. The creation of metaverse content is not fully leveraged by the current authoring tools and the existing techniques supporting user interaction with the Metaverse. In the past two decades, the human-computer interaction community has attempted to alleviate the ease of user interaction in diversified virtual environments. Nonetheless, usability gaps still exist, resulting in low efficiency and user frustration \cite{Human-shitty-Interact}. We see that such gaps will not be overcome if we purely rely on investigating user behaviours with alternative interfaces and interaction techniques, especially since the tasks inside virtual 3D spaces grow more complicated. 

Third, creating large objects, e.g., a dragon floating in mid-air, requires a relatively spatial environment. Users unavoidably encounter a lot of distal operations between the user position and the virtual creation. It is worth mentioning that users are prone to errors during such distal operations. A prior work \cite{Batmaz2019DoHD} provides evidence that users with headsets achieve lower pointing accuracy to a distal target. Considering such complicated operations in content creation, typical metaverse users can not immediately create objects except those already in the asset library. In other words, metaverse users have no appropriate approaches to unleash the full potential of creating content in the endless canvas of the Metaverse. Instead, they hire professionals to draw and mould virtual instances on traditional desktops. For virtual space owners, a team of professionals, e.g., unity developers, may spend days or weeks creating virtual environments.
Further change requests (e.g., adding a new 3D model) for such environments may take additional hours or days. Without time or skills, general users can only experience the contents being built by virtual space owners. As shown in Figure \ref{fig:issue}, this rigid circumstance is analogous to the `read mode' in Web 1.0. Creating unique metaverse content has become highly inconvenient and demanding. We will likely face the circumstance of `Web 1.0' in 3D virtual worlds, with some features inherited from Web 2.0, such as making new texts and uploading photos.

%The purpose of this article

To alleviate the barriers mentioned above, this article argues for using AI-generated content (AIGCs) in both content generation in the metaverse and AI-mediated user interaction in the metaverse. The article has a vision that the GPT-alike model can trigger content singularity in the Metaverse and assist interaction between human users and virtual objects in the Metaverse. 
Before we move on to the main discussion, we provide some background information regarding the Metaverse and AIGCs, as follows.

\begin{figure*}[hpt]
    \centering
    \includegraphics[width=15cm]{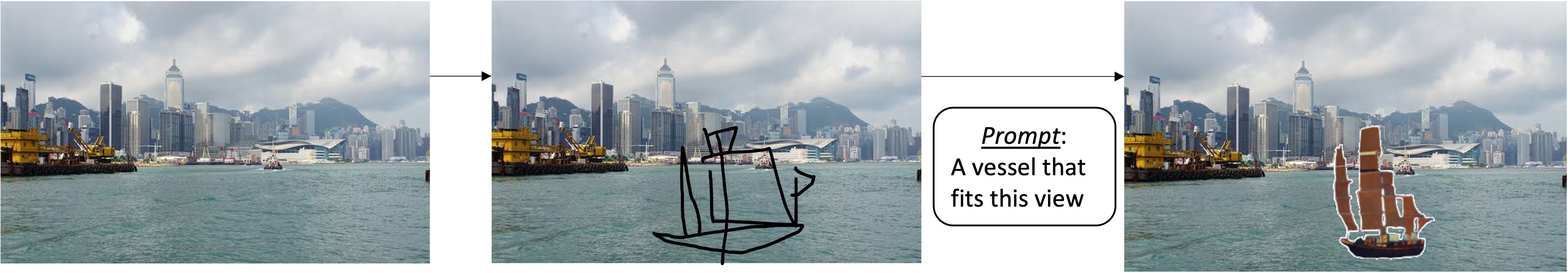}
    \caption{Generating a vessel that fits the context of Victoria Habour, Hong Kong. As a result, a junk boat appears: Original view (Left), Sketching (middle), the generated vessel on top of the physical world (Right). }
    \label{fig:vessel}
\end{figure*}

\textbf{Metaverse}: The Metaverse refers to the NEXT Internet featured with diversified virtual spaces and immersive experiences~\cite{Lee2022WhatIT}. Similar to existing cyberspace, we can regard the Metaverse as a gigantic application that simultaneously accommodates diverse types of countless users. The application comprises computer-mediated worlds under the Extended Reality (XR) spectrum and emerging derivatives like Diminished Reality (DR). Ideally, users will create content and engage in activities surrounding such content. Multitudinous underlying technologies serve as the backbone of the Metaverse, including AI, IoT, mobile networks, edge and cloud servers, etc. Among the technologies, we can view AI as the fuel to support the automation of various tasks and content creation. 
Our discussion in this article goes beyond the well-known applications, including creating avatars, virtual buildings, virtual computer characters and 3D objects, automatic digital twins, and personalized content presentation \cite{lee2021needs}.

\textbf{AI-Generated Content (AIGC)}: Apart from the analytical AI focusing on traditional problems like classification, AIGC can leverage high-dimensional data, such as text, images, audio, and video, to generate new content. For instance, OpenAI announces its conversational agent, ChatGPT~\cite{zhang2023ChatGPT}, of which the latest GPT-3 and GPT-4 can create texts and images, respectively. Moreover, the generated content can support the generation of metaverse objects, such as speech for in-game agents, 3D objects, artist artefacts and background scenes in many virtual worlds. The most popular techniques, including GANs, Diffusion models, and transformer architectures, support the challenging context-to-content task. It is important to note that generative AI and AIGC differ subtly \cite{Zhang2023ACS}. AIGC focuses on content production problems, whereas generative AI analyzes the underlying technological underpinnings that facilitate the development of multiple AIGC activities.

\section{Content Singularity}

The most widely used metaverse applications have appeared in industrial applications in the past two decades~\cite{industrial-AR}. The firms have the resources to build up proprietary systems and prepare the content of their interested domain. The work content drives the adoption of AR/VR applications in industrial sectors, with the following two examples. First, labour at warehouse docks and assembly lines can obtain helpful information (e.g., the next step) through the lens of AR \cite{AR-Assembly}. Second, personnel at elderly caring centres can nurture compassion from perspective-taking scenarios of virtual reality (VR) \cite{Paananen2022FromDM}. 
Content is one of the incentives, and end-users achieve enhanced abilities or knowledge, perhaps resulting in better productivity.

% then talk about 3D content in the Metaverse
As we discussed the main three barriers in \textit{Retrospect}, users have limited ability and resources to create unique content in the Metaverse. The general users can only draw some simple yet rough sketch to indicate an object in Extended Reality. 
Nonetheless, such expressiveness is insufficient for daily communication or on-site discussion for specific work tasks. We may expect the content on AR devices to be no worse than what we have in Web 2.0. To alleviate the issue, AIGCs can play an indispensable role in lowering the barriers and democratizing content creation.  
Figure~\ref{fig:vessel} illustrates a potential scenario where users can effectively create content in virtual-physical environments. 
For instance, a user with an AR device is situated in a tourist spot and attempts to show the iconic vessels to explain the cultural heritage of Hong Kong's Victoria Harbour. 
First, the AIGC model can understand the user's situation and context through sensors on the AR device, for instance, depth cameras. 
Second, the users can make a dirty and speedy sketch to indicate the shape and position of the generated object. In addition, the prompt which contains the user's description, i.e., a prompt of `a vessel fits this view', is sent to the AIGC model through methods like speech recognition. It is important to note that our speech always involves `this' or `that' to indicate a particular scene and object. The AIGC model can employ the user's situation and context in such a scenario. 
Finally, a junk boat appears in the Victoria Habour through the lens of AR. 

Singularity can refer to a point in time or a condition in which something undergoes a significant and irreversible milestone, depending on the context of such changes. Also, it is frequently used in technology and artificial intelligence (AI) to describe the hypothetical moment when robots or AI transcend human intellect and become self-improving or perhaps independent \cite{Prescott2013TheAS}. This notion is also known as technological singularity or AI singularity. It is a contentious issue to the Metaverse when AIGCs are widely adopted by end users. We believe the occurrence of AI-generated content might have far-reaching consequences for cyberspace. 
%The metaverse is missing content, content in web 2.0 can make a social world. 
Next, the concept of content singularity refers to the belief that we are reaching a time when there will be abundant virtual material available on the Internet that people will consume as their daily routine. 
This is owing to the demand for immersive cyberspace and related technological ability, perhaps AIGCs, to pave the path towards the exponential proliferation of virtual 3D content. This is similar to the social network in which people contribute and consumer content.

Since the launch of ChatGPT\footnote{\url{https://openai.com/blog/chatgpt}}, pioneering prototypes shed light on the daily uses of GPT-driven intelligence on AR wearables, such as generating simple 3D contents using WebAR (a.frame) by entering prompts\footnote{\url{https://www.youtube.com/watch?v=J6bSCVaXoDs&ab_channel=ARMRXR}} and providing suggested answers for conversations during datings and job interviews\footnote{\url{https://twitter.com/bryanhpchiang/status/1639830383616487426?cxt=HHwWhMDTtfbC7MEtAAAA}}. 
These examples go beyond the industrial scenarios, implying that AIGC-driven conversational interfaces can open new opportunities for enriching virtual-physical blended environments~\cite{zhou2023unleasing}. 
Generative AI models can recognise the user context using the sensors on mobile devices (e.g., cameras on AR headsets or smartphones) to generate appropriate objects according to given prompts. In this decade, general users will treat generative AI models as utilities like water, electricity, and mobile network. Meanwhile, the metaverse is an endless container to display AI-generated content so users can read and interact with the AI utility mid-air. Users can make speech prompts to generative AI models to create characters, objects, backdrop scenes, buildings, and even audio feedback or speeches in virtual 3D environments. 
These content generations should not pose any hurdle or technical difficulties to the general users. It will be as simple as posting a new photo on Instagram, typing a tweet on Twitter, or uploading a new video on Tiktok. The lowered barrier will encourage people to create content, and more content consumers will follow, eventually leading to a metaverse community. 
%(9) AIGC -> economy (virtual) values. 
In addition, rewarding schemes should be established when the content singularity arrives to sustain the content creation ecosystem. AIs, the data owners behind them, and users become the primary enablers and principal actors, respectively. The way of splitting the reward among them is still unknown, and ongoing debates will continue. 

%(7) Diminished Reality (DR) - the necessity of Redefining AR, and Ethics (naked body of a girl), 
Generative AI models are obviously drivers of content generation. But we should not neglect its potential of removing contents, primarily physical counterparts, through the lens of XR devices, also known as Diminished Reality (DR). It is important to note that the naive approach of overlaying digital content on top of the physical world may hurt the user experience. 
%(2) Matchness with the 3D or physical worlds? In 2D worlds, we don't care about the physical counterpart. 
A virtual instance may not match the environmental context, and it may be necessary to change the context to show better perceptions when the metaverse application strongly relates to daily functions. We may accept a virtual Pokémon appearing on top of a physical rubbish bin. However, we feel weird when a virtual table overlaps the physical table being disposed of. 
Therefore, AIGCs may serve as a critical step of DR to smoothen the subsequent addition of digital overlays (AR). In this sense, the demands of AIGCs will penetrate throughout the entire process of metaverse content generation. More importantly, the diminished items should comply with the user's safety or ethical issues. Hiddening a warning sign may put the users in danger. Also, putting off a person's clothes may show inappropriate content, i.e., a naked body. It is essential to reinforce regulation and compliance when generative AI models are widely adopted in the content-generation pipeline. 

%Also, it will be impossible for anyone to absorb it all. 
On the other hand, content singularity can also refer to the challenges of information overload in a virtual-physical blended environment, in which people are assaulted with so much information that it is impossible to digest and make sense of it all \cite{A2W-LAM2021}. 
% we first talk about knowledge and then context awarenesss. 2D text as the most applicable example. 
The sheer volume of online information, including text, photos, videos, and music, is already daunting and rapidly increasing. As such, the virtual-physical blended environment may cause a lot of disturbance to users if we neglect such exponential proliferation of 3D content.

%(17) Helping FOV and display content, too small, too big? Context-aware again?
Information or knowledge in the tangible world can indeed be thought limitless, whereas augmentation within the relatively limited field of view of headsets is complex. Consequently, we must optimise the presentation of digital content. Typically, metaverse users with a naive approach to virtual content delivery will experience information inundation, thereby requiring additional time to consume the augmentation. Context awareness, such as the users, environments, and social dynamics, is a prominent strategy for managing the information display. The AIGCs at the periphery, with the assistance of recommendation systems, can interpret user context and provide the most pertinent augmentation~\cite{A2W-LAM2021}.

%(20) uniqueness - repeat its importance and calls for some new shits
Although we foresee a rise in content volume when AIGCs are fully engaged as a utility in the Metaverse, two significant issues should be addressed.
First, content uniqueness raises concerns about the quality and relevancy of the material provided. With so much material accessible, users are finding it increasingly difficult to identify what they seek and discern between high-quality and low-quality content. To address the issues of content singularity, additional research studies should have been made to create new tools and methodologies that will assist users in filtering, prioritizing, and personalizing the material they consume. Current solutions in Web 2.0 include search engines, recommendation algorithms, and content curation tools. Yet, the issue of content singularity remains a complicated and continuing one that will undoubtedly need further innovation and adaptation as the volume and diversity of digital information increase in the Metaverse.

%(16) Explainability - current conversational interfaces 
Second, contemporary conversational interfaces have long been criticized for lacking transparency as a `black box' \cite{Arrieta2019ExplainableAI}. In other words, conversational AIs do not show a complete list of their ability, while the general users usually have no clues about what the AI can achieve. Significantly, users with low AI literacy cannot quickly master the interaction with GPT-like AI agents through a conversational interface. Exploring the perfect fit between the generative AI models and the XR environment is necessary. For instance, the AI models can suggest some potential actions to the users by putting digital overlays on top of the user's surroundings. As such, the user can understand the AI's ability and will not make ineffective enquiries or wasted interactions with the generative AI model. 
%(12b) Blackbox of AIGC - user has no clue what can be done with the AIGC-driven model. Also, What cannot be done? 
In addition, more intuitive clues should be prepared, according to the user context, to inform the user about `what cannot be done' with a generative AI model. 

\begin{figure*}[hpt]
    \centering
    \includegraphics[width=15.5cm]{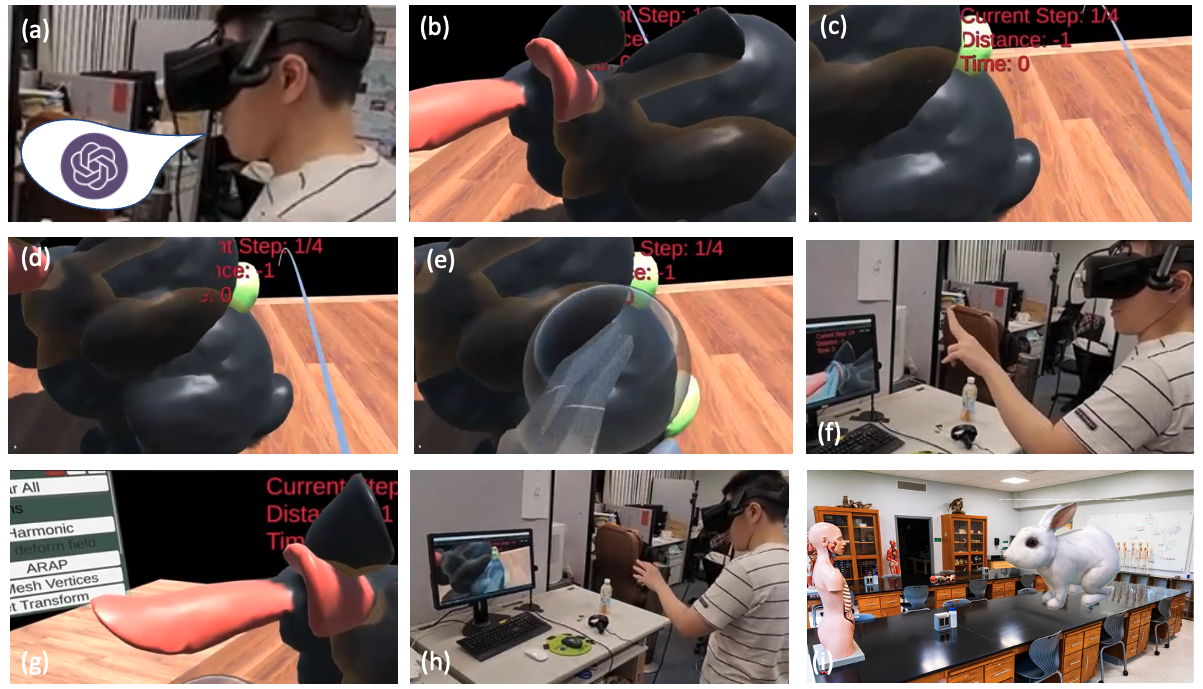}
    \caption{An example pipeline of content creation and human-metaverse interaction supported by AIGCs: (a) brainstorming with conversational agents (collecting requirements simultaneously); (b) auto-generation of the contents; (c) start manual edition but huge pointing errors exist; (d) following (c), AI-assisted pointing for selecting vertex; (e) following (d), AI-assisted vertex editing; (f) manual editing of subtle parts; (g) AI-assigned panel and user interaction on the virtual objects; (h) user reviews of the objects while AIGCs attempt to understand the user perceptions; (I) content sharing, e.g., educational purpose in a classroom. 
    Photos are extracted and modified from \cite{3DeformR} for illustration purposes.}
    \label{fig:rabbit}
\end{figure*}

\section{Human-Metaverse Interaction}
Besides generating virtual content, AIGC can be considered an assistive tool for user interaction in the metaverse. 
From other users' perspectives, a user's movements and interaction with virtual objects can be a part of the content in virtual worlds. The difficulties of controlling an avatar's movements and interacting with virtual objects can negatively impact an individual's workload and the group's perceptions of a metaverse application. For example, a group awaits an individual to finish a task, causing frustration.

Before discussing how the prompt should be extended in the Metaverse for easier interaction between users and metaverse instances, some fundamentals are considered in human-computer interaction (HCI) and prompt engineering \cite{Liu2021DesignGF}. 
%(5) How we can set the prompt engineering (prompt <-> hci <-> context) --> leading to context-to-content, for the sake of lowering the user efforts and improving ease-of-use
Prompts have different concerns in HCI and NLP. From the HCI perspective, Effective prompts are clear, concise, and intuitive. Users have to design prompts for an interactive system, and users' workloads exist to take specific actions or provide relevant input. Once the user's needs and goals have been identified, the next step is to craft effective prompts that guide the user towards achieving those goals. And the AI-generated results provide users with the information they need to take action in a particular context. Therefore, prompt engineering is an essential aspect of designing interactive systems that are easy to use and achieve high levels of user satisfaction. 

Prompt engineering, in NLP and particularly LLMs, refers to the methods for how communicating with LLM to steer their behavior for desired outcomes.  %involves creating clear, concise, and easy-to-understand prompts to enhance the user experience and improve the metaverse's overall usability. 
The traditional chatbot (e.g., ChatGPT) considers primarily text prompts. In contrast, the prompts from the metaverse users can become more diverse by considering both the context as discussed above and multiple user modalities, including gaze, body movements, and psychological and physiological factors. In addition, perhaps employing certain personalization techniques, prompts should be tested and refined iteratively to ensure that they effectively guide LLMs towards the desired output. 
As such, metaverse-centric prompt engineering requires a new understanding of the user's needs and goals, as well as their cognitive abilities and limitations. This information can be gathered through user testing, A/B testing, user surveys and usability testing in many virtual worlds. 

The prompt design can be extended to the subtle interaction between virtual objects and users. VR sculpting is a popular application where users can freely mould their virtual objects in virtual spaces. The usability of VR, inaccurate pointing to vertex, becomes a hurdle \cite{3DeformR}. It is still far away from being the main tool of creativity due to its efficiency. A hybrid model can be considered: generative AI models like LLMs can first generate a model of 3D content, and then we customize the model with manual editing in VR. In this sense, an important issue arises -- we cannot get rid of manual operations with virtual instances. AIGCs, in the future, should assist human users in virtual tasks that inherit the nature of complexity and clumsiness, under hardware constraints, such as limited Field-of-view (FOV).
AIGCs can parse the user actions in virtual environments, for instance, limb movements and gazes towards a virtual object, to provide appropriate work done from the manual editing. As such, AIGCs can serve as assistants for metaverse users.
%(14) Content Creation Tools as an example - we can sketch for creating content, but we rely on the AIGC models to provide the details. Can we extend this concept to our interaction? My limb movement, for instance, we can leverage the existing work on visual cues. VR sculpting 
It is important to note that AI-assisted tasks always happen on the daily ubiquitous devices, i.e., smartphones. A prevalent example of 2D UIs is typing text on soft keyboards. Users tap on the key repetitively and make typos if triggered by adjacent keys. Such an erroneous task can be assisted by auto-correction. Users can tap the mistyped word and select the correct spelling from the suggested words. To achieve this, an AI model learns the words in the English dictionary and then understands user habits by recording the user's word choice. 

%Furthermore, 
%Arguments start with mobile text inputs....
%dictionary - habits - prediction word of text.
%Errorenous operations on 2D interfaces.

Typing on a soft keyboard is a good example of an AI-assisted task. In virtual environments, the interaction tasks, including dragging an object to a precise position and editing an object of irregular shapes, can be challenging to the users. AIGCs open opportunities to help human users accomplish the task. Nonetheless, the typing tasks on soft keyboards can be manageable because the dictionary is a reasonable search space. In contrast, AIGC-driven assistance can encounter a much larger room. In the editing task, a user can first select a vertex at a rabbit's tail. The next action can be changing the vertex property and then moving to another vertex. The next vertex can happen on the head, bottom, etc. With the current technology, predicting the user's following action at a highly accurate rate is very unlikely. However, if available, AIGCs may leverage the prior users' behaviours from a dataset containing user interaction footprint, and accordingly recommend several `next' edits to facilitate the process. Eventually, the user can choose one of them and accomplish the task without huge burdens. 

\begin{figure*}[hpt]
    \centering
    \includegraphics[width=15cm]{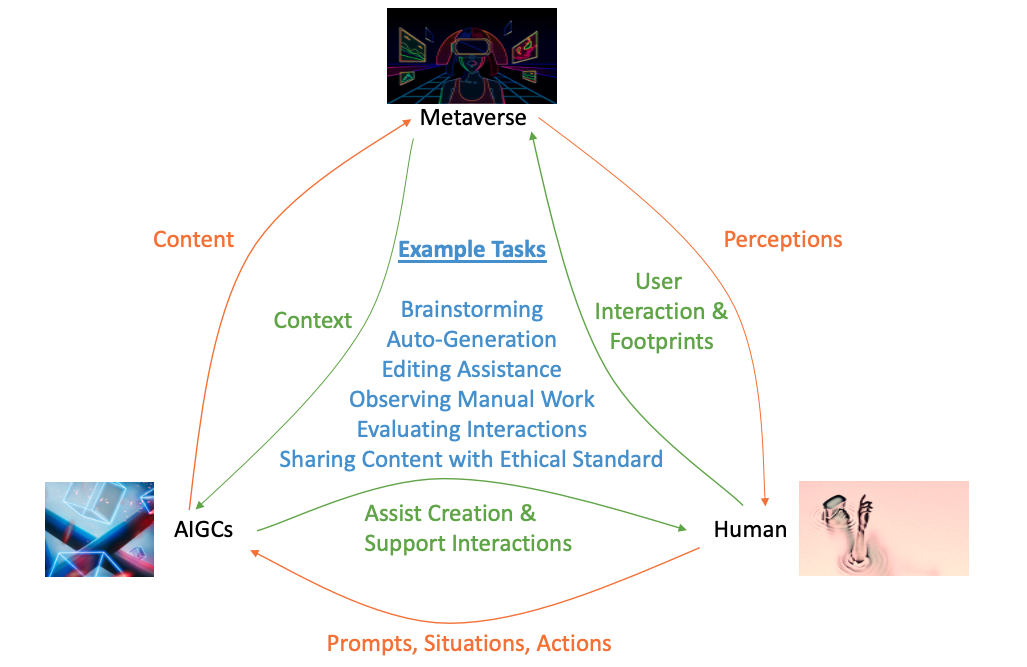}
    \caption{AIGM framework showing the relationship between human users, AIGCs and virtual-physical cyberspace (i.e., the Metaverse).}
    \label{fig:AIGM}
\end{figure*}

In a broader sense, diversified items exist in many virtual worlds, and a virtual item can have many possible relationships with another. As such, the user interaction with AIGCs' predictions becomes complicated. For instance, I pick up an apple and then lift a tool to cut it. Other possible actions include putting down the apple, grabbing an orange, etc. It is also important to note that building an ontology for unlimited items in the Metaverse is nearly impossible. 
One potential tactic is to leverage the user's in-situ actions. Generative AI models can read the user's head and hand movements to predict the user's interested regions and, thus, upcoming activities.
Ideally, a user may give a rough pointing location to a particular item. Then, Generative AI models can make personalized and in-situ suggestions for the user's subsequent interactions with virtual objects, with sufficient visualization to ensure intuitiveness. 
We believe that the above examples are only the tip of the iceberg but sufficient to illustrate the necessity of re-engineering the ways of making metaverse-ready prompts for Generative AI models. 
%(11) Using text input as the argument. Beyond Accuracy  

%draw figure again -- edit a precise object - the rabbit with AIGC? a model-assisted by AIGC and a model solely by human's will?

%(3) more diversified interaction types and task than ever before, foot movement, hand movement, object movement, to name but few. In the past, we only need to focus on the movement of the mouse point, how it clicks on an object (menu or window, icons, etc) Now, we have to pay attention to many aspects of user interaction -- perhaps increased cognitive leads, demands, and frustration. 

%(15) User movement as the new context of new input. User inputs as the new prompt engineering?

%(22) not necessarily a human avatar, what about flying as a flock of birds. What about transmitting our gestures like some fluids? 

% (1) Task changes: 2D --> 3D time/effort/perception?

% (4) which button the user should choose, what are the visual cues, and what are the reasonable choices of UI tasks? Do we have the good judgement or UI/UX tools for XR interfaces? We can only focus on one item once at a time. We still cannot get rid of 2D- inherited interface?

% (6) cyberspace becomes more complex (UI-wise) than the counterpart of 2D worlds. we may draw a figure to illustrate this.

%(18) obvious the current chatgpt services knowledge provision, as a form of intelligence augmentation, or advanced intelligence. But can this be used to augment the user ability in XR?

Then, there is the issue of how natural people will feel in the metaverse environments built, or in some cases hallucinated, with AIGCs. Urban designers and architects are now looking into what factors of our ordinary environments matter most when attempting to translate the environments into digital ones, beyond the 3D replication. Here, issues such as \textit{subjective presence} (comfort, feeling, safety, senses) or active involvement (activities taking place, other people's presence), in addition to the traditionally considered \textit{structural aspects} (colour, furnishing, scale, textures), will play a pivotal role in how the metaverse experience will feel like for its users (see, e.g., \cite{paananen2021investigating}). And so the questions to solve will include such as to what degree will we want generative AI to be able to spawn experiences that feel safe, or should the spaces more closely reflect the world as we know it, outside the metaverse, where different even adjacent spaces will have a lot of different perceived human characteristics to them. 

The technical capability of AIGCs only opens a landscape of generating metaverse content, regardless of adding backdrops (AR) or removing objects causing strong emotions (DR). But we know very little about user aspects once AIGCs can be scaled. As the metaverse moves beyond the sole digital interfaces, i.e., 2D UIs, the AIGC can be embedded in the physical worlds and alter the user's situated environments for fulfilling users' \textit{subjective presence} that can be abstract. It can vary greatly due to the user's beliefs (norms, customs, ego, and so on) and their environment. A machine may not truly interpret the meaning of `calm', especially if multiple subjective presences are underlying, e.g., `safe and calm'. A user makes a simple prompt of `calm' to an AIGC model. Consequently, the results are unsatisfactory, as the user does not make effective prompts, for example, adding words like `meditation, wellness and sleep' if the users are inside a bedroom. 
It is worth noting users with headsets may expect quick and accurate feedback, instead of requesting the generative AI models to revise the content with multiple iterations. 
In addition, \textit{subjective presence} does not limit to a single user. Multiple users will interact with metaverse content in a shared space, potentially causing co-perception and communication issues.
Generating the right content at the right time leads to a challenging problem more than technical aspects. AIGC in the Metaverse will lead to a novel niche of understanding the dynamics among metaverse content, physical space, and users.

%The above discussion only considers how a single user percepts the physical worlds with virtual content with AIGCs. Nonetheless, metaverse content will share among users, and  sharing 
%An environment hosts multiple users sharing the same virtual-physical space. 

%The user actions and their dynamic with the situated environments can become the 

\section{Towards AIGM Framework}

We argue AIGM is a must -- should we aim to unleash all of the latent potentials in the metaverse concept. This is, regardless of who is the leading developer, the metaverse must be built for humans, and as humans, everything we do is embodied in the space around us \cite{paananen2021investigating}. 
%User-Environment-Utility
The leading developers do not have the authority to arrange what content we should have on the Next Internet, as we have seen in the Metaverse of 2022, in which the virtual spaces are office-like environments. We usually spend eight work hours at the physical office, and it is insane to spend the rest of 8 hours in the virtual office. Ironically, except for the standard item given in asset libraries, we don't have the right to decorate such office space with our unique creations. 
It is eventually the user's call for the popular trend in the Metaverse. When Google's Image searches were conducted since Q3 2021, it was evident that the creators had always defined the metaverse with blue, dark, and purple colours. 
We believe the trend of popular content is ever-changing. Driven by the vital role of AIGCs in democratizing content creation, everyone in the Metaverse can decide, (co-)create, and promote their unique content. 
To scale up the use of AIGCs, we propose a framework for an AI-Generated Metaverse (AIGM) that depicts the relationships among AIGCs, virtual-physical blended worlds, and human users, see Figure~\ref{fig:AIGM}. AIGC is the fuel to spark the content singularity, and Metaverse content is expected to surround everyone like the atmosphere. 
This creates an entire creation pipeline in which AIGCs are the key actors. First, the users can talk to generative AI models to obtain inspiration during human-AI conversations (Human-AI collaboration). 
Consequently, generative AI models provide the very first edition of the generated content (AI-Generation). It then supports subtle editing during content creation (AI-Assistance). Some precise details can be done manually (Human users); if necessary, multiple users can be involved in the task (Multi-user collaboration). 
In addition, it is important to note that AIGCs can assign properties of how the users and virtual instances will interact, e.g., through a tap on a panel, and accordingly, AIGC-driven evaluations will perform to understand the user performance and their cognitive loads \cite{Hu2023ExploringTD}.
Eventually, content sharing and the corresponding user interaction can be backed by AIGCs.

\section{Concluding Remarks}
During a deceleration of global metaverse development, the author contends that AIGCs can be a critical facilitator for the Metaverse. This article shares some perspectives and visions of when AIGCs meet the Metaverse. 
Our discussion started with a look back at the key flaws of metaverse applications in 2022. We also highlight the fundamental difficulties the metaverse encountered. Accordingly, we examine how AIGCs will speed up metaverse development from a user standpoint. The article eventually speculates on future possibilities that combine the Metaverse with AIGCs. We call for a conceptual framework of AIGM that facilitates content singularity and human-metaverse interaction in the AIGC era. We also hope to provide a more expansive discussion within the HCI and AI communities.

%\input{sections/background}
% \input{sections/architecture}
% \input{sections/cv}
% \input{sections/network}
% \input{sections/interaction}
% \input{sections/privacy}
% \input{sections/evaluation}
% \input{sections/implementation}

% In a number of respects, our creative ambition exceeds what is yet technically feasible. Even with multi-billion dollar investments from the world's greatest digital giants, every other week, customers approach us with ideas for fantastic experiences on paper, but the technology is just not there yet to make them a reality. Because of the promise of a socially-driven virtual world, there is such a big push. 

% When businesses dipped their toes into the experimentation pool in 2022, they brought playful experiences to numerous platforms, giving their communities something to do. There is value in a shared environment that can exist across devices, but only forward-thinking businesses are now investing in it. 

% The fascinating difficulty I see is that the metaverse is already fundamentally split among the many firms establishing their own platforms, like Meta, Roblox, Niantic, and Epic Games, to mention a few. Each community has its own metaverse in which to devote time and resources. Hence, when businesses approach a single platform, they only tap into a small portion of the metaverse. As we construct the basis of what the metaverse will become, this raises the question: is there a method for marketers to publish experiences across different metaverse platforms?

\bibliographystyle{IEEEtran}
\balance
\bibliography{references,privacy}

\end{document}